# Gate-tunable hot electron extraction in a two-dimensional semiconductor heterojunction

*Chenran Xu[1,2,8], Chen Xu[1,2,8], Jichen Zhou[1,2], Zhexu Shan[1,2], Wenjian Su[1,2], Wenbing Li[1,2], Xingqi Xu[1,2], Kenji Watanabe[3], Takashi Taniguchi[4], Shiyao Zhu[1,2,5,6,7], Da-Wei Wang[1,2] and Yanhao Tang[1,2]\**

[1] School of Physics, Zhejiang University, Hangzhou 310027, China

[2] Zhejiang Key Laboratory of Micro-Nano Quantum Chips and Quantum Control, Zhejiang University, Hangzhou 310027, China

[3] Research Center for Electronic and Optical Materials, National Institute for Materials Science, 1-1 Namiki, 305-0044 Tsukuba, Japan

[4] Research Center for Materials Nanoarchitectonics, National Institute for Materials Science, 1-1 Namiki, 305-0044 Tsukuba, Japan

[5] State Key Laboratory for Extreme Photonics and Instrumentation, Zhejiang University, Hangzhou 310027, China

[6] College of Optical Science and Engineering, Zhejiang University, Hangzhou 310027, China

[7] Hefei National Laboratory, Hefei, China

[8] These authors contributed equally

*Email: yanhaotc@zju.edu.cn

ABSTRACT

Hot carrier solar cells (HCSCs), harvesting excess energy of the hot carriers generated by above-band-gap photoexcitation, is crucial for pushing the solar cell efficiency beyond the Shockley–Queisser limit, which is challenging to realize mainly due to fast hot-carrier cooling. By performing transient reflectance spectroscopy in a $MoSe_2$/hBN/$WS_2$ junction, we demonstrate the gate-tunable harvest of hot electrons from $MoSe_2$ to $WS_2$. By spectrally distinguishing hot-electron extraction from lattice temperature increase, we find that electrostatically doped electrons in $MoSe_2$ can boost hot-electron extraction density ($n_{ET}$) by factor up to several tens. Such enhancement arises from interaction between hot excitons and doped electrons, which converts the excess energy of hot excitons to excitations of the Fermi sea and hence generates hot electrons. Moreover, $n_{ET}$ can be further enhanced by reducing the conduction band offset with external electric field. Our results provide in-depth insights into design of HCSCs with electrostatic strategies.

TEXT

When a semiconductor absorbs photons with energy above its band gap, hot carriers with excess kinetic energy are generated. One solution of harvesting such energy, termed as hot carrier solar cells (HCSCs)[1], is to extract hot carriers to the higher-energy band of energy selective contacts (ESCs) before cooling, as illustrated in Figure 1c. The perfect HCSCs can increase the efficiency up to 66%[2], about twice the Shockley–Queisser (SQ) limit[2,3]. However, this approach is extremely challenging[4], since most of the excess energy is quickly lost to the lattice by optical phonon emission typically within several picoseconds (ps). Although efforts have been devoted to slowing down the phonon-assisted cooling process, such as exploiting hot-phonon bottleneck effect[5,6] and trapping hot carriers in higher-energy side valleys[7], no HCSCs beyond the SQ limit have been realized yet.

Two-dimensional (2D) materials act as a promising platform for the hot-electron extraction owing to the strong carrier-carrier scattering, which can compete with the electron-phonon scattering due to the reduced screening effect[6]. In addition, 2D heterostructures can be fabricated by choosing plentiful types of material and possess the high electrical-tunability[8], which could offer abundant opportunities in optimizing the efficiency of hot-electron extraction. For example, graphene can efficiently convert absorbed photon energy to hot carriers[9], which can be extracted to the higher-energy band of adjacent semiconductor[10,11]. However, zero band gap of graphene is not favored for photovoltaic applications, since most of hot carriers will relax to the Dirac point within several ps[9]. In this regard, semiconductor transition metal dichalcogenides (TMD) with band gap of ~eV and strong light-matter interaction[12–14] are better candidates for HCSCs. Ultrafast dynamics studies[8,15,16] have revealed fast interface charge transfer within 100 femtoseconds (fs) in TMD heterostructures, in which photoexcited carriers transfer to the lower-energy band in the other layer and all excess energy is lost. The extraction of hot carriers in TMD to a nearby higher-energy band[17] still remains a challenge, which is vital for realizing HCSCs.

In this study, we fabricate a dual-gated MoSe$_2$/hBN/WS$_2$ heterostructure wherein MoSe$_2$ and WS$_2$ play the roles of the photoabsorber and the ESC of HCSCs, respectively, and perform transient reflectance (TR) spectroscopy. We demonstrate highly tunable hot-electron extraction dynamics from the conduction band of MoSe$_2$ to the higher-energy conduction band of WS$_2$. Specifically, we develop a rigorous method of extracting $n_{ET}$ from the TR spectrum, as well as photoinduced increment of lattice temperature. Upon electron doping MoSe$_2$, $n_{ET}$ shows a giant enhancement with a factor of several tens that arises from efficient scatterings between hot excitons and doped electrons, and scales linearly within a wide range of applied electric field. Our results highlight the exciton-electron interaction as a novel strategy for harvesting the excess energy of hot carriers in TMD materials.

Figure 1a shows a dual-gated device schematic. MoSe$_2$ and WS$_2$ monolayers are separated by a thin hBN layer (~1 nm), and grounded via a graphite electrode; top and bottom gate voltages ($V_t$ and $V_b$) are applied. See device fabrication and characterization in the Supporting Information (SI), as well as Figure S1 and S2. As shown in Figure 1b, the heterostructure has an intrinsic type-I band alignment[18,19], where the band gap of MoSe$_2$ is fully encompassed within that of WS$_2$. This is verified by the doping dependence of steady reflectance contrast (RC) spectrum (Figure 1d). Upon both electron and hole doping, the A-exciton resonance of MoSe$_2$ ($A_M$) is suppressed and the Fermi polarons[20–24] ($A_M^-$ and $A_M^+$) form, but little changes for the A-exciton resonance of WS$_2$ ($A_W$). The intrinsic conduction and valence band offsets ($\Delta_C$ and $\Delta_V$) are respectively determined to be about 170 meV and 250 meV (Figure S3), consistent with previous studies[18,19]. Doping is modulated by $V_D = V_t + 1.15V_b$, where the coefficient of 1.15 mainly counts for different thicknesses of top and bottom hBN dielectrics; for a given $V_D$, the band offsets are modulated by $V_E = V_t - 1.15V_b$ (see the SI for details). Unless otherwise specified, all optical measurements are performed at a sample temperature of 10 K.

To explore the hot-electron extraction, we perform TR spectroscopy with the scheme illustrated in Figure 1b. Analogous to the principle of HCSCs (Figure 1c), MoSe$_2$ works as a photoabsorber with tunable electron doping density $n_e$; WS$_2$ works as a broadband ESC with tunable band offset $\Delta_C$. A pump pulse with photon energy about 1.746 eV is to generate hot excitons in MoSe$_2$ without exciting WS$_2$ (Figure 1e); a broadband pulse is to probe optical resonances of WS$_2$ and thus monitor hot-electron transfer from MoSe$_2$ to WS$_2$; the time delay ($\tau$) between pump and probe pulses is varied (see the SI for details of the ultrafast setup). Hot-hole transfer is not favored, due to the much larger $\Delta_V$. To be noted, WS$_2$ is kept electrostatically neutral for all ultrafast measurements.

Let's first take a glance at TR spectra of optical resonances of WS$_2$ with MoSe$_2$ electron doped. Figure 2a shows the contour plot of the TR spectrum at varying $\tau$, from which several spectral linecuts are extracted (Figure 2b). The TR signal $\Delta R/R$ is defined as $(R_{on} - R_{off})/R_{off}$, where $R_{on}$ and $R_{off}$ are the reflected probe spectra with and without pump, respectively. The TR spectrum shows appearance of negative Fermi polaron resonance ($A_W^-$) of WS$_2$ around 2.03 eV after several ps, evidencing transferred hot electrons in WS$_2$. For the more pronounced TR feature around $A_W$ resonance (~2.06 eV), it shows a complicated temporal evolution: at $\tau$ =0.8 ps, the spectral weight of $\Delta R/R < 0$ is much larger than that of $\Delta R/R > 0$; from 0.8 to 10 ps, the spectral weights of $\Delta R/R < 0$ and $\Delta R/R > 0$ gradually become comparable; after 10 ps, the spectral profile slightly changes. In the following, we will focus on $A_W$ resonance for its large signal-to-noise ratio.

This complicated evolution indicates multiple effects following the optical excitation, and here we consider three effects—transferred hot electrons, lattice temperature increment, and enhanced surrounding screening, the last of which likely arises from mobile photoexcited carriers in MoSe$_2$. We identify the TR spectral profiles of $A_W$ resonance induced by these effects independently, based on the steady RC spectrum modulated by: electrostatically doping WS$_2$ that corresponds to the electron-transfer TR profile (Figure S4); increasing sample temperature that corresponds to the temperature-increase TR profile (Figure S5); electrostatically doping MoSe$_2$ that corresponds to the screening TR profile (Figure S6). As shown in Figure S7, we can see that the temperature-increase and screening TR profiles are almost the same, which are different from the electron-transfer TR profile. We expect that temperature increase has a larger contribution to the TR signal than that of the screening, because photoexcited excitons in MoSe$_2$ are dipole-like neutral quasiparticles that should have weak remote screening effect on the hBN-separated WS$_2$. See the SI for more discussion about the three effects.

We use the two TR profiles of electron transfer and temperature increase as the bases to fit the experimental TR spectrum, referred as the two-profile analysis, which can give $n_{ET}$ and lattice temperature increment $\Delta T^*$. Figure 2c and d show the TR spectra at $\tau = 1.2$ and 40 ps, respectively, in spite of distinct shapes, both of which can be well fit by the two-profile analysis. We extract $n_{ET}$ and $\Delta T^*$ as a function of $\tau$ shown in Figure 2e and f, respectively, except initial delays ($\leq 0.4$ ps) that suffer from contaminating effects of coherent light-matter interaction (see fittings at other time delays in Figure S8). The $n_{ET}$ and $\Delta T^*$ dynamics are further fit by a two-exponential function (dashed) that gives the rise and decay time constants. $n_{ET}$ shows a rise time constant of 2.6 ps and a decay time constant of 138 ps. The rise dynamics of $n_{ET}$ is the convolution of the electronic thermalization (e.g., via exciton-electron scatterings) that generate hot electrons and the hot-electron tunneling from MoSe$_2$ to WS$_2$. Here, we attribute the rise dynamics of $n_{ET}$ to the electronic thermalization in MoSe$_2$, as a result of faster hot-electron tunneling (<1 ps, see more discussion later); the decay dynamics of $n_{ET}$ is attributed to the electron back-transfer from WS$_2$ to MoSe$_2$, driven by the conduction band offset, which is consistent with another ultrafast study of hBN-separated TMD heterostructures[25]. The magnitude of $\Delta T^*$ is also in line with previous reports[26,27].

With the two-profile analysis, we explore hot-electron extraction dynamics with varying electron doping in MoSe$_2$, as shown in Figure 3a. The maxima of $n_{ET}$ dynamics, denoted as $n_{ET,m}$, are shown in Figure 3b. Upon MoSe$_2$ electron doped, $n_{ET,m}$ gets vastly enhanced by a factor up to ~30, with an onset behavior. Moreover, the rise dynamics also shows an onset behavior, which suddenly slows down at low $n_e$ and then becomes faster with further increasing $n_e$ (Figure 3c). The largest $n_{ET,m}$ corresponds to the hot-electron extraction efficiency about 5%, defined as $n_{ET,m}/n_{ex}$, where $n_{ex}$ is the initial photoexcited hot exciton density in MoSe$_2$. The corresponding $\Delta T^*$ dynamics is shown in Figure S9.

To gain more insights about doping-enhanced hot-electron extraction, we study ultrafast dynamics of $A_M$ resonance of MoSe$_2$ with varying electron doping in MoSe$_2$. Surprisingly, the rise timescale (empty circles) of $A_M$ resonance shows onset doping behavior similar to that of $n_{ET}$, as shown in Figure 3c and Figure S10. Upon electron doping MoSe$_2$, both rise timescales of A$_M$ and $n_{ET}$ increase up to about 6 ps; as electron doping further increases, both timescales decrease to about 1 ps. The rise dynamics of A$_M$ resonance reflects the electronic thermalization in MoSe$_2$, during which hot excitons interact with the Fermi sea and thus hot electrons are generated. Since the dynamics of $n_{ET}$ is the temporal convolution of the electronic thermalization and the hot-electron tunneling, the similarity of the rise timescales evidences that, the rise dynamics of $n_{ET}$ can be mainly attributed to the electronic thermalization, and the hot-electron tunneling itself should be much faster (<1 ps that corresponds to the shortest rise timescale of $n_{ET}$ dynamics at $V_D$=6.45 V). Similar fast hot-electron tunneling has been observed in WS$_2$/graphene heterostructures[10,28]. Hence, we propose a doping-assisted mechanism illustrated in Figure 3e,f: Through the scatterings between excitons and electrons in MoSe$_2$, the excess energy of hot excitons is converted to the excitations of the Fermi sea, which generate hot electrons with enough kinetic energy that can transfer to the higher-energy conduction band of WS$_2$; higher electron doping leads to more frequent scatterings (i.e., faster electronic thermalization), which generate more hot electrons and lead to faster rise dynamics of $n_{ET}$. Once the electronic quasi-equilibrium is achieved with the formation of Fermi polaron, no more hot electrons get generated and the hot-electron extraction terminates. For neutral MoSe$_2$ (Figure 3d), the scatterings among excitons can also generate hot electrons that can transfer to WS$_2$, likely via Auger recombination—the recombination of one exciton leads to the ionization of the other exciton. It should be noted that, it is hard to extract hot-electron transfer from the $A_M$ dynamics of MoSe$_2$, since which involves multiple other effects, such as interaction between photo-excited excitons and doped electrons, electron-hole recombination, and temperature change.

We further explore the power dependence of hot-electron extraction with electron doped and neutral MoSe$_2$, as shown in Figure 4a and b, respectively. For doped MoSe$_2$, when the pump fluence increases, $n_{ET}$ shows overall increase and faster rise dynamics. In Figure 4c, $n_{ET,m}$ exhibits a sublinear power dependence that gives enhanced hot-electron extraction efficiency for a lower pump fluence (Figure 4d). This is a favored character for photovoltaic applications under weak illumination intensity, such as sunlight, in contrast to the hot-phonon bottle neck effect that requires high illumination intensity[5,6]. Moreover, for neutral MoSe$_2$, $n_{ET,m}$ shows a linear-like power dependence with the extrapolation deviating from the origin (Figure 4c), which gives increased hot-electron extraction efficiency for a higher pump fluence (Figure 4d). See Figure S11 for the corresponding $\Delta T^*$ dynamics.

The power dependence of the hot-electron extraction efficiency can be understood in terms of the aforementioned mechanism (Figure 3). As the pump fluence increases, the amount of the excess energy to be dissipated increases. However, for a given electron-doping density, the dissipation capacity of the Fermi sea is limited by the scattering rate between hot excitons and electrons. As the excess energy exceeds the capacity of the Fermi sea, other dissipation channels such as the electron-phonon scatterings become more involved, which will lead to a reduced hot-electron extraction efficiency at high pump fluence. In contrast, for neutral MoSe$_2$, hot electrons arise from Auger recombination of excitons, in which increasing initial exciton density leads to more efficient generation of hot electrons.

Lastly, for shedding light on the impact of the conduction band offset, we explore the electric field dependence of the hot-electron extraction with doped and neutral MoSe$_2$, as shown in Figure 5a and b, respectively. For both doped and neutral MoSe$_2$, as $V_E$ increases and thus $\Delta_C$ decreases, $n_{ET}$ shows overall increase and slower decay dynamics. This can be understood as follows. Reducing $\Delta_C$ gives rise to more hot electrons that have enough kinetic energy to transfer to WS$_2$, and less driving force for the electron transfer from WS$_2$ back to MoSe$_2$. Worthy to note, the electron transfer back to MoSe$_2$ will only slightly modify the decay dynamics of A$_M$, instead of leading to any rise, owing to other overwhelming processes in MoSe$_2$, such as the recombination of electron-hole pairs. Moreover, as shown in Figure 5c,d, the maximum of $n_{ET}$ shows linear dependence within a wide range of $V_E$, indicated by dashed lines. This is distinct from the Fowler-Nordheim tunneling for the devices with thick hBN barrier[29–32] wherein the tunneling current is highly nonlinear with the bias voltage[29], which means that, in this work, hot electrons in MoSe$_2$ directly tunnel to WS$_2$, without travelling to the bands of the inset hBN. See Figure S12 for the corresponding $\Delta T^*$ dynamics.

In this work, we demonstrate gate-tunable hot-electron extraction, as well as a rigorous method of obtaining both transferred electron density and lattice temperature increment, which demonstrate a doping-assisted mechanism in harvesting excess energy of hot carriers in TMD materials. Although the current hot-electron extraction efficiency is only a few percent, it can be further enhanced by reducing the thickness of inset hBN (e.g., the tunneling probability of 1L hBN is about one order larger than that of 3L hBN[33]), using a multilayer $MoSe_2$ with indirect band gap[12,13,34] that can reduce radiative recombination losses, and aligning the crystal orientation of both TMDs layers for minimizing momentum mismatch.

FIGURES

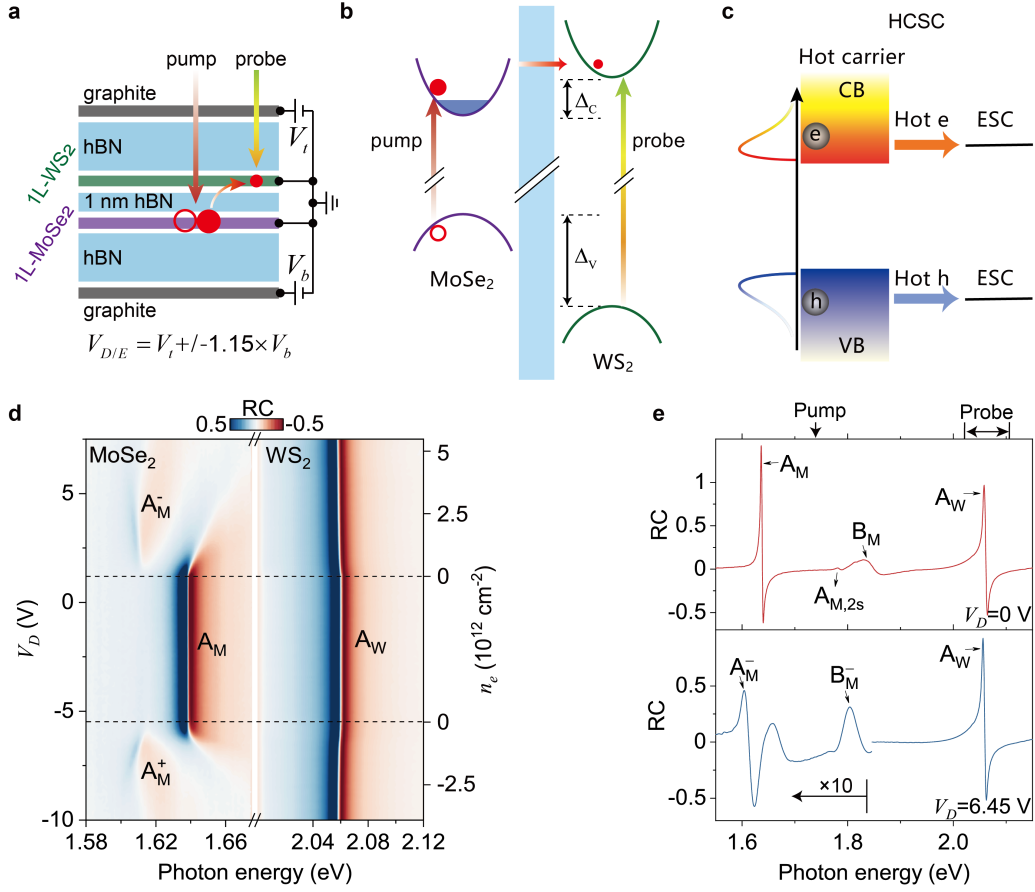

**Figure 1.** Schematics of the device, hot carrier photovoltaics, and pump-probe scheme. (a) Schematic of a dual-gated WS$_2$/hBN/MoSe$_2$ heterostructure. WS$_2$ and MoSe$_2$ monolayers are grounded; top and bottom gate voltages ($V_t$ and $V_b$) are applied. (b) For the pump-probe scheme, only MoSe$_2$ is optically excited, and optical resonances of WS$_2$ are probed for monitoring hot-electron extraction dynamics, with varying doping and band offsets. (c) The concept of HCSCs. Photoexcited hot electrons and holes are collected by the electron ESC (above the conduction band minimum) and the hole ESC (below the valence band maximum), respectively. (d) Contour plot of the doping dependence of steady RC spectrum with $V_E = 0$ V. (e) Two RC spectra with neutral and electron doped MoSe$_2$, extracted from (d). $A_M$, $A_M^{-/+}$, $B_M$, $B_M^-$ represent A-exciton, A-exciton-derived Fermi polaron, B-exciton, and B-exciton-derived Fermi polaron resonances of MoSe$_2$, respectively. Similar label conventions are applied to reflectance resonances of WS$_2$.

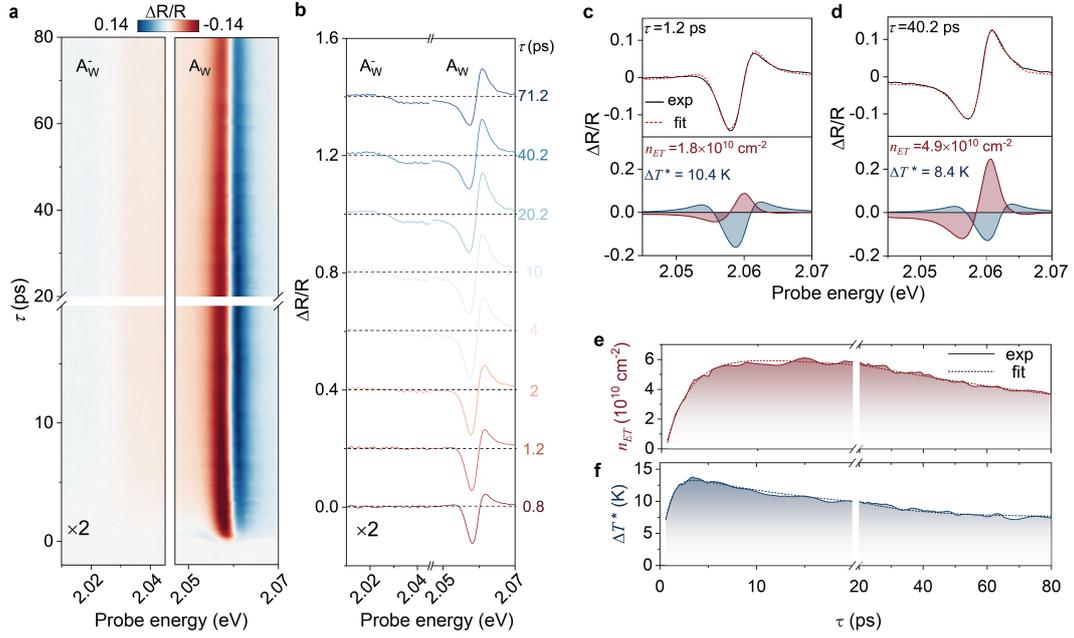

**Figure 2.** The hot-electron extraction and lattice temperature increase distinguished in transient reflectance spectra. (a) Contour plot of transient reflectance (TR) spectrum as a function of pump-probe delay $\tau$ with probe energy covering $A_W$ and $A_W^-$ resonances of $WS_2$. With $V_D = 2.6$ V and $V_E = 0$ V, $n_e$ in $MoSe_2$ and $\Delta_C$ are about $1.3 \times 10^{12}$ cm$^{-2}$ and 120 meV, respectively. The pump fluence of 63 $\mu$J/cm$^2$ generates an initial exciton density $n_{ex}$ about $2.5 \times 10^{12}$ cm$^{-2}$, which is determined by multilayer interference calculation (see the SI). (b) Spectral linecuts at various $\tau$, extracted from (a). The curves are vertically displaced for clarity, with dashed lines denoting $\Delta R/R=0$. (c, d) The two-profile analysis of the TR spectra at $\tau=1.2$ and 40.2 ps. The upper panel shows the experimental data and fitting; the lower panel shows the two TR profiles used in the fitting, revealing $n_{ET}$ and $\Delta T^*$. (e, f) The dynamics of $n_{ET}$ and $\Delta T^*$, fitted with a two-exponential function (dashed).

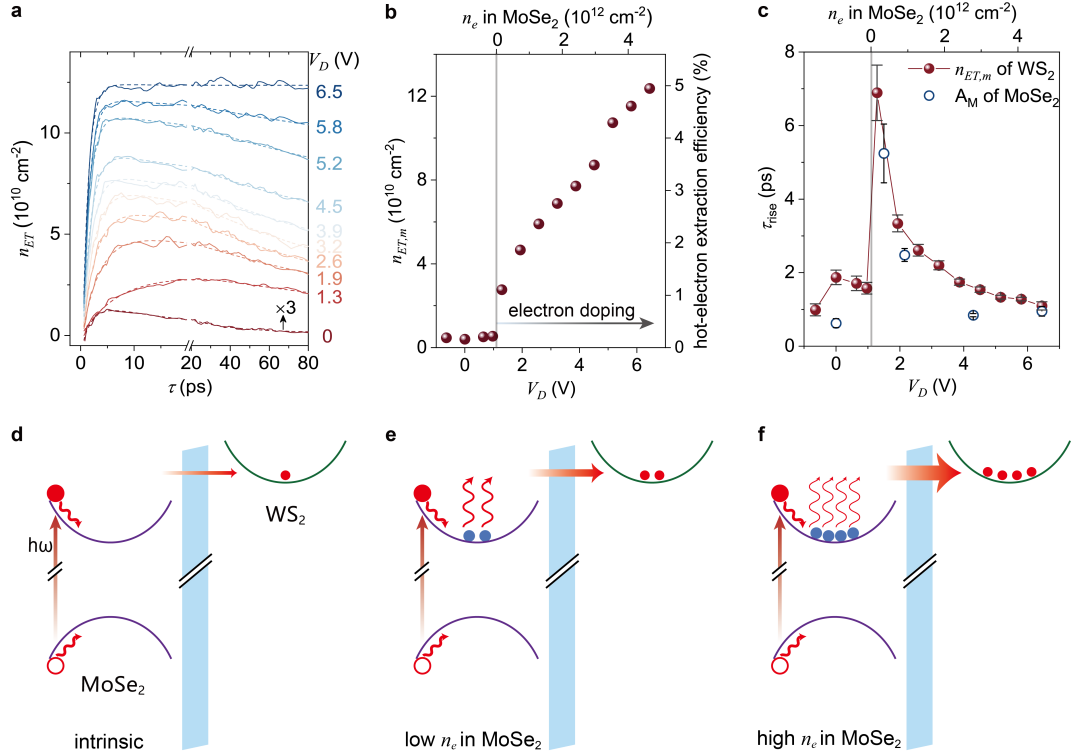

**Figure 3.** Enhanced hot-electron extraction upon electron doping MoSe$_2$. (a) The $n_{ET}$ dynamics with $n_e$ varying from 0 to 4.6×10$^{12}$ cm$^{-2}$ in MoSe$_2$, corresponding to $V_D$ varying from 0 V to 6.45 V. The initial $n_{ex}$ is about 2.5×10$^{12}$ cm$^{-2}$, corresponding to pump fluence about 63 μJ/cm$^2$. $V_E$ is kept at 0 V. The dynamics is fitted with a two-exponential function (dashed). (b) The maximum of hot-electron extraction density $n_{ET,m}$ and the corresponding hot-electron extraction efficiency as a function of $V_D$. The vertical gray line indicates the initial of electron doping. (c) The doping-dependent rise time constant of $n_{ET}$ as a function of $V_D$, compared with the rise time constant of A$_M$ resonance (Figure S10). (d-f) The schematic of hot-electron extraction mechanism for neutral and doped MoSe$_2$. The scatterings between hot excitons and doped electrons (e,f) can generate hot electrons more efficiently than the scatterings among excitons (d). See main text for detailed discussion.

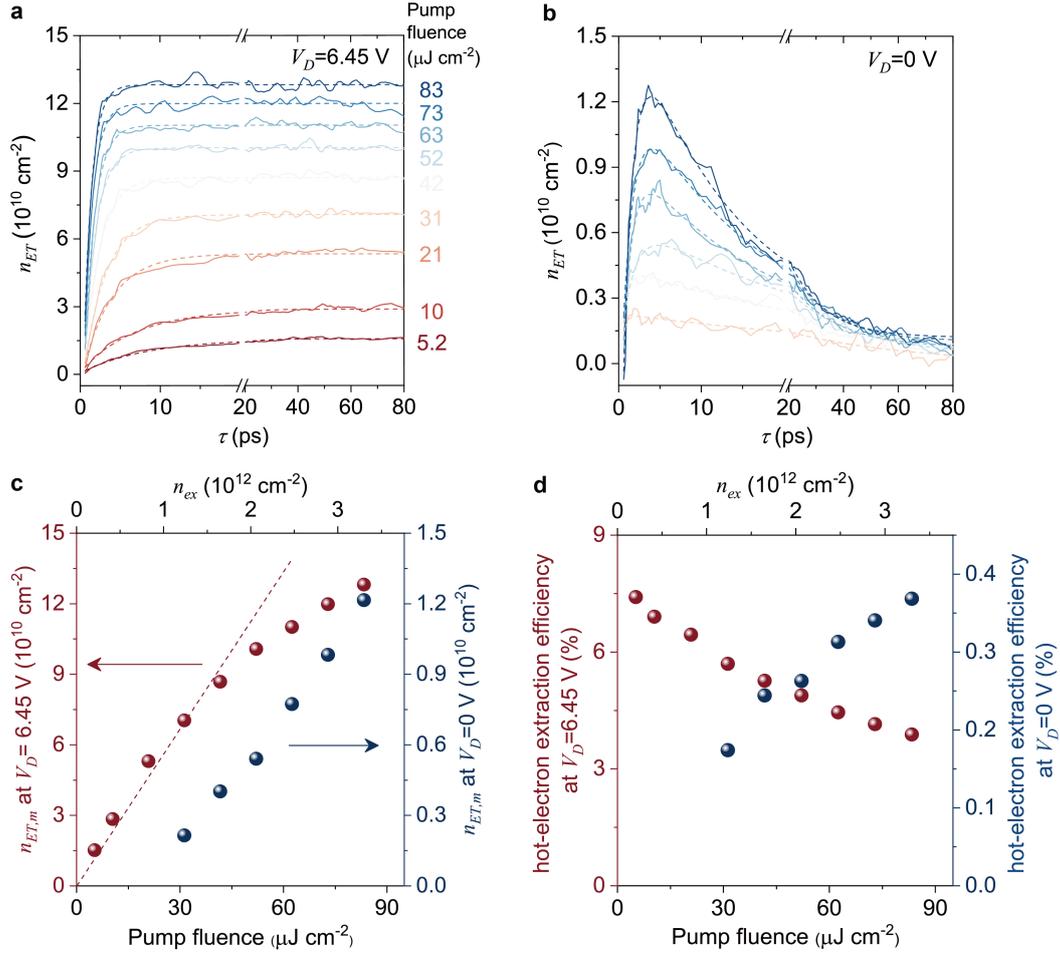

**Figure 4.** The power dependence of hot-electron extraction with electron doped and neutral MoSe$_2$. (a, b) The $n_{ET}$ dynamics with pump fluence varying from 5.2 to 83 μJ/cm$^2$, corresponding to $n_{ex}$ from about 0.2 to 3.3×10$^{12}$ cm$^{-2}$. $V_D$ is set to 6.45 V and 0 V, corresponding to electron doped MoSe$_2$ with $n_e$ about 4.6×10$^{12}$ cm$^{-2}$ and neutral MoSe$_2$, respectively. $V_E$ is kept at 0 V. The dynamics is fitted with a two-exponential function (dashed). (c, d) The power dependence of $n_{ET,m}$ (c) and hot-electron extraction efficiency (d) for the two doping scenarios. The straight dashed line is to highlight sub-linear behavior.

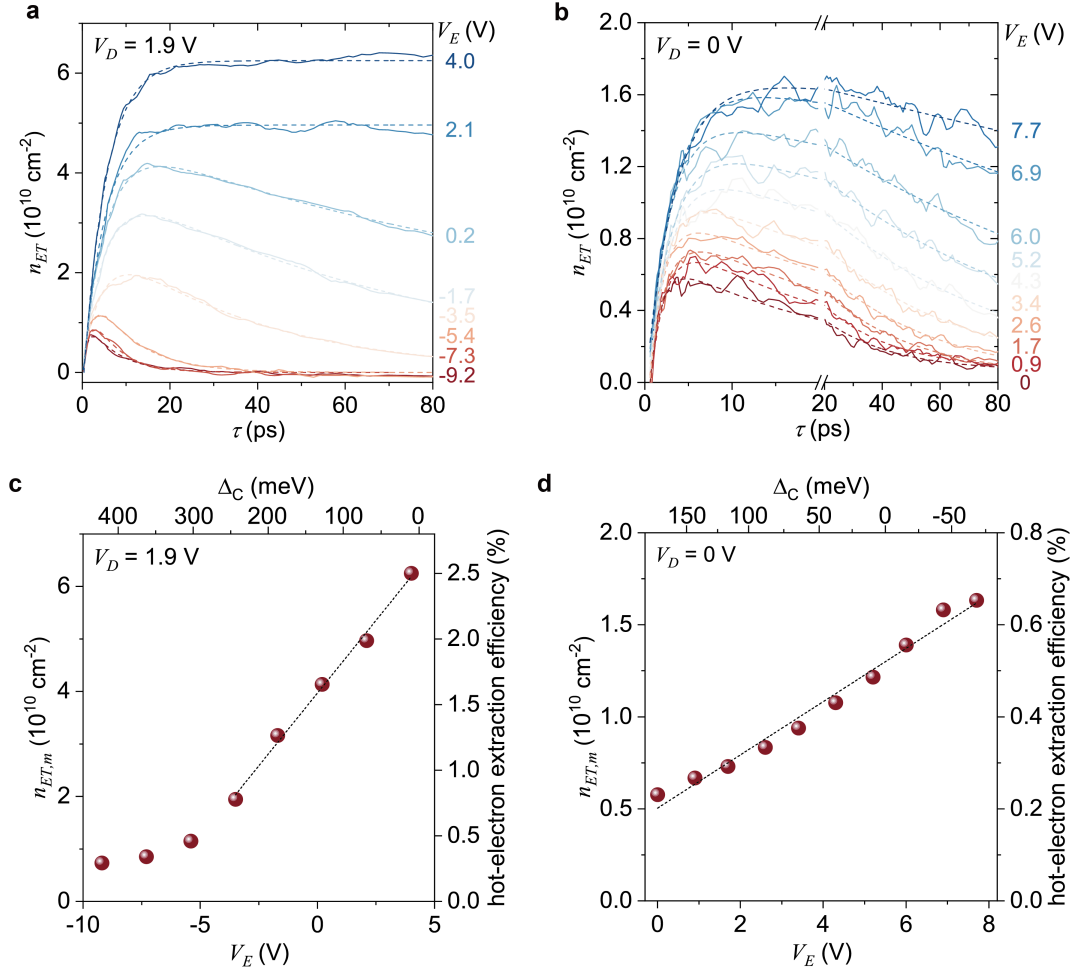

**Figure 5.** The electric field dependence of hot-electron extraction with electron doped and neutral MoSe$_2$. (a, b) The $n_{ET}$ dynamics with varying $V_E$. $V_D$ is set to 1.9 V and 0 V, corresponding to electron doped MoSe$_2$ with $n_e$ about $0.7 \times 10^{12}$ cm$^{-2}$ (a) and neutral MoSe$_2$ (b), respectively. The dynamics is fitted with a two-exponential function. (c, d) The electric field dependences of $n_{ET,m}$ and the corresponding hot-electron extraction efficiency for the two doping scenarios. Dashed lines are eye-guidance to the linear behavior.

ASSOCIATED CONTENT

**Supporting Information**

Device fabrication and characterization, the steady RC spectroscopy, the transient reflectance spectroscopy, electrostatic calculation of $n_e$ and $\Delta_C$, estimation of initial $n_{ex}$, identifying the TR profiles of the $A_W$ resonance induced by various effects, the two-profile analysis of the TR spectrum, $\Delta T^*$ dynamics with varying doping in MoSe$_2$, the TR spectrum around the $A_M$ resonance upon electron doping MoSe$_2$, the pump-fluence dependence of $\Delta T^*$ dynamics, the electric-field dependence of $\Delta T^*$ dynamics, and discussion of hot-electron extraction at ambient condition

AUTHOR INFORMATION

**Corresponding Author**

Yanhao Tang: yanhaotc@zju.edu.cn

**Author contributions:** Y.T. designed the scientific objectives; Y.T., D.W. and S.Z. supervised the project; C.X. built up the ultrafast setup and performed TR spectroscopy under Y.T.'s supervision, with assistance from X.X.; C.X. and Y.T. analyzed the data; Chen Xu and J.Z. fabricated the devices, with assistance from Z.S., W.S. and W.L.; K.W. and T.T. grew the hBN crystals; Y.T. and C.X. wrote the manuscript with the input from all the authors. C.X. and Chen Xu contributed equally to this work.

**Competing interests**

The authors declare no competing interests.

ACKNOWLEDGMENT

This work was supported by the National Key R&D Program of China (grant nos. 2022YFA1405400 and 2022YFA1402400), the National Natural Science Foundation of China (grant nos. 12274365, 11934011, and 12325412), Zhejiang Provincial Natural Science Foundation of China (grant no. LR24A040001), and Open project of Key Laboratory of Artificial Structures and Quantum Control (Ministry of Education) of Shanghai Jiao Tong University. K.W. and T.T. acknowledge support from the JSPS KAKENHI (Grant Numbers 21H05233 and 23H02052) and World Premier International Research Center Initiative (WPI), MEXT, Japan. In addition, we appreciate the device fabrication support from the ZJU Micro-Nano Fabrication Center in Zhejiang University.

# Supporting Information for

# "Gate-tunable hot electron extraction in a two-dimensional semiconductor heterojunction"

*Chenran Xu[1,2,8], Chen Xu[1,2,8], Jichen Zhou[1,2], Zhexu Shan[1,2], Wenjian Su[1,2], Wenbing Li[1,2], Xingqi Xu[1,2], Kenji Watanabe[3], Takashi Taniguchi[4], Shiyao Zhu[1,2,5,6,7], Da-Wei Wang[1,2] and Yanhao Tang[1,2]\**

[1] School of Physics, Zhejiang University, Hangzhou 310027, China

[2] Zhejiang Key Laboratory of Micro-Nano Quantum Chips and Quantum Control, Zhejiang University, Hangzhou 310027, China

[3] Research Center for Electronic and Optical Materials, National Institute for Materials Science, 1-1 Namiki, 305-0044 Tsukuba, Japan

[4] Research Center for Materials Nanoarchitectonics, National Institute for Materials Science, 1-1 Namiki, 305-0044 Tsukuba, Japan

[5] State Key Laboratory for Extreme Photonics and Instrumentation, Zhejiang University, Hangzhou 310027, China

[6] College of Optical Science and Engineering, Zhejiang University, Hangzhou 310027, China

[7] Hefei National Laboratory, Hefei, China

[8] These authors contributed equally

\*Email: yanhaotc@zju.edu.cn

# Content

# 1. Device fabrication and characterization

## 1.1 Fabrication of the MoSe$_2$/hBN/WS$_2$ device

The dual-gate MoSe$_2$/hBN/WS$_2$ heterostructure was fabricated by the dry transfer method[1]. Specifically, atomically thin flakes were exfoliated from bulk crystals onto 280 nm SiO$_2$/Si substrates, and identified by an optical microscope. All the flakes were picked up sequentially with a polymer stamp—a polycarbonate (PC) layer on a polypropylene-carbonate-coated polydimethylsiloxane (PDMS) substrate, and released onto a sapphire chip with prepatterned electrodes. The whole transfer process was performed in a nitrogen-filled glovebox. See the optical image of the device in Figure S1.

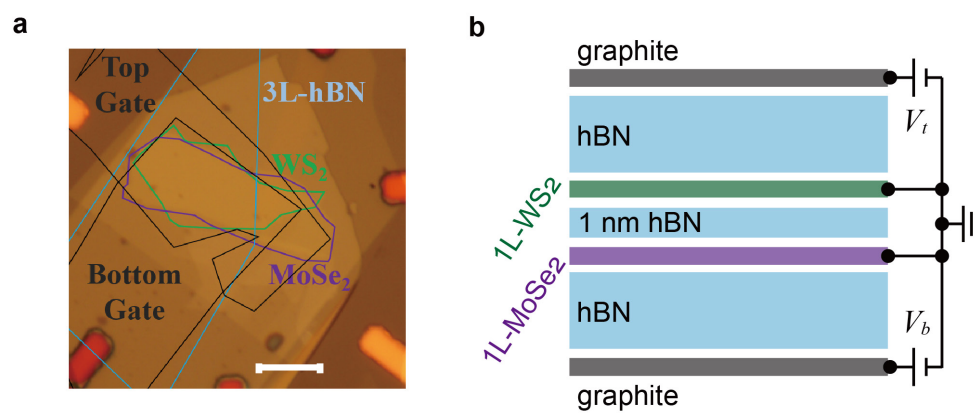

**Figure S1.** The optical image of the device. (a) The green, purple and blue lines denote the sample edges of WS$_2$, MoSe$_2$ and inset hBN, respectively. The black lines indicate the top and bottom graphite gates. The scale bar is 10 $\mu$m. (b) Schematic of the device.

## 1.2 Characterization of the MoSe$_2$/hBN/WS$_2$ device

Figure S2a shows the PL spectrum at the heterostructure in the charge-neutral condition, exhibiting the A$_M$ and A$_W$ resonances with linewidths of 1.7 meV and 2.7 meV, respectively. Figure S2b shows the Raman spectrum from monolayer WS$_2$, monolayer MoSe$_2$, and the overlapped region of both TMD monolayers, the last of which exhibits A$_{1g}$ mode from MoSe$_2$ (242 cm$^{-1}$) and E$_{2g}$ mode from WS$_2$ (356 cm$^{-1}$). Figure S2d displays the reflectance contrast mapping of the heterostructure. The gray circle marks the uniform area where the optical measurements were performed. Figure S2e-j show the AFM measurements of the thicknesses of top hBN, bottom hBN and inset hBN.

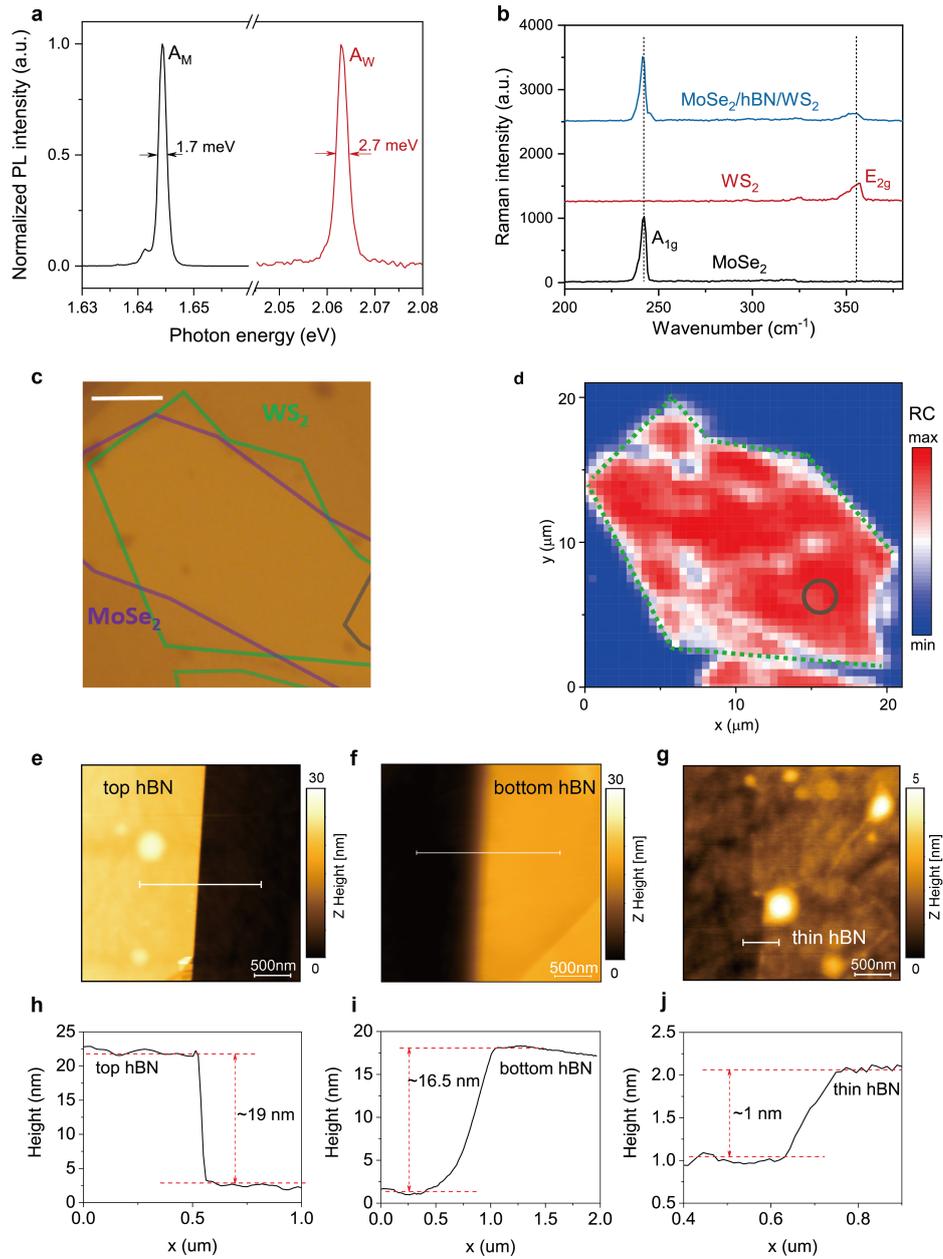

**Figure S2.** Characterization of the MoSe$_2$/hBN/WS$_2$ device. (a) Photoluminescence (PL) spectrum from the device with 532 nm excitation. (b) Raman spectra from region of WS$_2$, MoSe$_2$ and heterostructure. (c) The optical microscopy image of the device, with the scale bar of 10 μm. (d) The reflectance contrast mapping with the size range the same as (c). The color bar represents the magnitude of the reflectance contrast of the A$_W$ resonance. The dotted line marks the outline of WS$_2$. (e-j) The atomic force microscope measurements of top hBN (e,h), bottom hBN (f,i) and inset hBN (g,j). The spectrum measurements (a,b,d) are performed at cryogenic temperature.

## 2. The steady RC spectroscopy

The output from a halogen lamp was used as the light source for RC measurements, which was collected with a single mode fiber, collimated with a 10× objective, and focused onto the sample by an objective (50×, NA=0.6). The reflected light was collected with the same 50× objective, and analyzed by a spectrometer (Horiba, iHR550) coupled with air-cooled charge-coupled-device camera. The RC spectrum was obtained by comparing the reflected light spectrum ($R_s$) of interest with a featureless spectrum ($R_r$) from the sample upon highly doping or the substrate, which was defined as $RC \equiv \frac{R_s - R_r}{R_r}$. The sample was mounted onto a cold finger of a closed-cycle cryostat (Cryo Industries of America) with base temperature of 10 K. The top and bottom gate voltages are controlled by two sourcemeters (Keithley 2400).

## 3. The transient reflectance spectroscopy

The transient reflectance (TR) spectroscopy was performed in the same reflectance geometry used in the steady RC measurements. The pump pulse was from the output of an optical parametric amplifier (OPA, Light Conversion, ORPHEUS-HP) driven by a 200-KHz ultrafast laser (1047 nm, Spirit 1040-30-HE, Spectra Physics). The probe pulse was a supercontinuum produced by focusing the fundamental light from the laser to a 3-mm sapphire crystal. The central wavelength of the pump was 710 nm (1.746 eV), and the pump fluence was adjusted by a polarizer and half-wave plate. The probe was filtered to a narrow spectral range of interest (e.g., 2.01 to 2.10 eV for detecting resonances of WS$_2$, 1.58 to 1.69 eV for detecting resonances of MoSe$_2$), for minimizing the fluence that is kept below 1 µJ/cm$^2$. The pump and probe beam sizes at the sample were 5.5 and 1.9 µm, respectively. The pump was chopped at a frequency of 481 Hz. The reflected probe was analyzed by a home-built spectrometer (1200 gr/mm grating; 100 mm focal length) coupled with a fast line scan camera (Hamamatsu C15821-2351). The overall temporal resolution is 600 fs. For each delay, the TR spectrum was acquired with exposure time of six seconds. The dynamics of the TR signal and extracted quantities (e.g., $n_{ET}$ and $\Delta T^*$) are analyzed by a two-exponential function, $A \cdot e^{-\tau/\tau_{rise}} + B \cdot e^{-\tau/\tau_{decay}} + C$. $A$, $B$, $C$, $\tau_{rise}$ and $\tau_{decay}$ are fitting parameters, the last two of which are the rise and decay time constants.

## 4. Electrostatic calculation of $n_e$ and $\Delta_C$

As shown in Figure S3f, $n_e$ in MoSe$_2$ can be obtained by the parallel-plate capacitor model, given by $n_e = \frac{\varepsilon \varepsilon_0 \Delta V_t}{e(d_t + \frac{\varepsilon}{\varepsilon_w} d_w + d_0)} + \frac{\varepsilon \varepsilon_0 \Delta V_b}{e d_b}$. Here, $d_t (\approx 19$ nm), $d_b (\approx 16.5$ nm), $d_0 (\approx 1$ nm), and $d_w (\approx 0.6$ nm) are the thicknesses of top hBN, bottom hBN, inset hBN, and monolayer WS$_2$, respectively, determined by either an atomic force microscope(AFM) or the contrast analysis of the optical image; $\varepsilon (\approx 3)$[2] and $\varepsilon_w (\approx 7)$[3] are the dielectric constants of hBN and WS$_2$; $\Delta V_t$ and $\Delta V_b$ are the top and bottom gate voltages relative to that of doping onset; $\varepsilon_0$ and $e$ are vacuum permittivity and elementary charge, respectively. $n_e$ is further approximately reduced to $\frac{\varepsilon\varepsilon_0}{ed_t}\Delta V_D$, where $\Delta V_D = \Delta V_t + 1.15\Delta V_b$.

$\Delta_C$ is modulated by the electric fields between MoSe2 and WS2, which is given by $\Delta_C = \Delta_C^i - ed_w E_w - ed_0 E_0$. $E_w$ and $E_0$ are the electric fields in monolayer WS2 and inset hBN (Figure S3f), respectively, which is related by $\varepsilon_w E_w = \varepsilon E_0$; $\Delta_C^i$ is the intrinsic $\Delta_C$ with both of $E_w$ and $E_0$ equal to 0 V/nm. For electron doped MoSe2 that is well grounded and can be treated as a metal, $E_0$ is determined by the top gate, which gives $E_0 = \frac{V_t}{d_t + d_0 + \frac{\varepsilon}{\varepsilon_w}d_w}$. Using $V_t = \frac{V_D + V_E}{2}$, $\Delta_C$ equals to $\Delta_C^i - e\eta\frac{V_D + V_E}{2}$ with $\eta = \frac{d_0 + \frac{\varepsilon}{\varepsilon_w}d_w}{d_t + d_0 + \frac{\varepsilon}{\varepsilon_w}d_w}$.

To experimentally determine $\Delta_C$ at various $V_D$ and $V_E$, we measured the RC spectrum of resonances of MoSe2 and WS2 as a function of $V_E$ with $V_D$ fixed at various values, as shown in Figure S3a-e. As $V_E$ increases to a given value ($V_E^*$), the $A_W$ resonance of WS2 gets suppressed which means that WS2 starts to get electron doped. At $V_E^*$, we have $\Delta_C^i - e\eta\frac{V_D + V_E^*}{2} \approx 0$ meV. In Figure S3g, we can see that the doping-dependent $V_E^*$ can be well fitted by a linear function. Thus, $\Delta_C$ is given as $e\eta\frac{V_E^* - V_E}{2}$, where $V_E^*$ is obtained by the linear fitting.

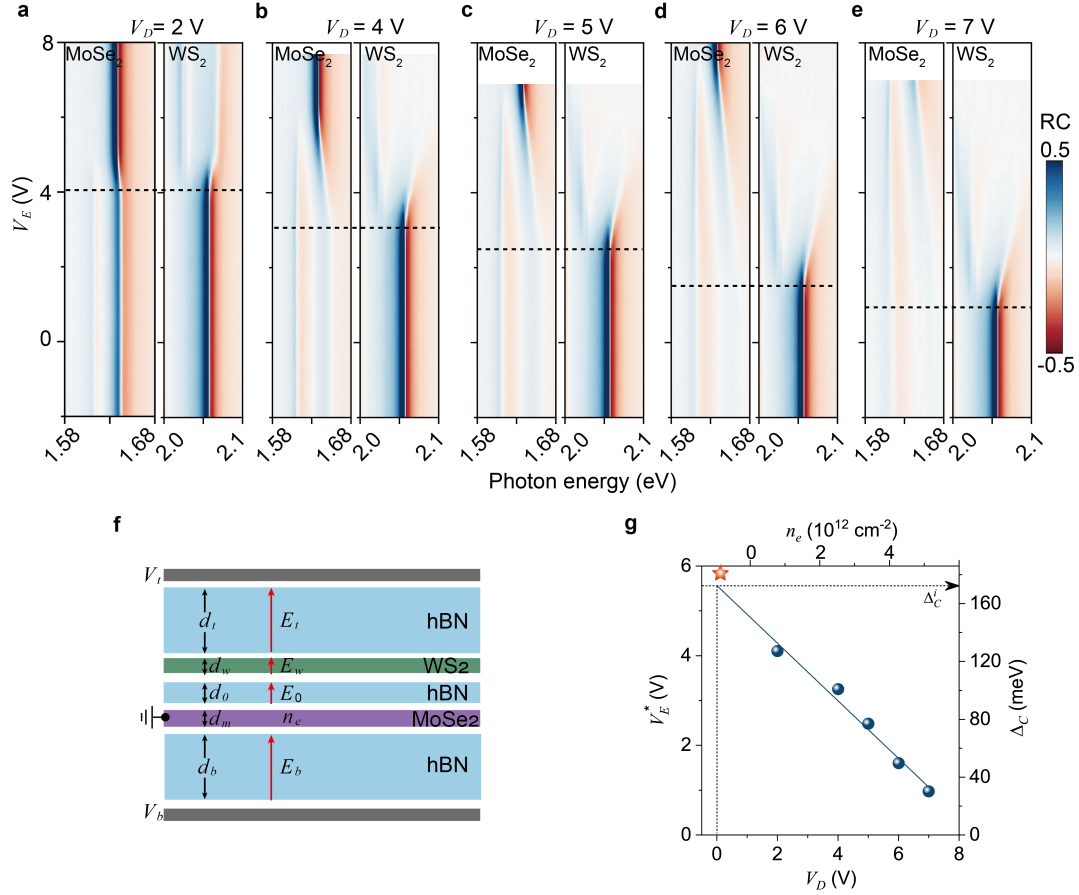

**Figure S3.** Determination of the conduction band offset between MoSe2 and WS2. (a-e) The contour plots of the RC spectrum as a function of $V_E$ at various $V_D$. The dashed line denotes the critical $V_E$, denoted as $V_E^*$, above which doped electrons start to move from MoSe2 to WS2. (f) The electric fields within the device. (g) The $V_E^*$ as a function of $V_D$. The straight line is a linear fitting, from which we can obtain the doping-dependent $\Delta_C$ with $V_E$=0 V, shown as the right y-axis.

## 5. Estimation of initial $n_{ex}$

For estimating $n_{ex}$, we first calculated the interference within the device with the transfer matrix method. By using the thickness of each layer that was obtained either by AFM or by the contrast analysis of the optical image, the pump intensity ($I$) in MoSe2 is related with that out of the device ($I_0$), by $I \approx 1.75 I_0$. Then, we estimated how many photons absorbed by monolayer MoSe2 at light frequency of $\omega$, by $\left(1 - e^{-\frac{2\kappa\omega d_m}{c}}\right) I$, and assumed that each photon generates one exciton. $c$ is the speed of light in vacuum, and $\kappa$($\approx$0.6 for 1.746 eV)[4] is the imaginary part of the refractive index of MoSe2. With the calculation above, about 1.1% of $I_0$ are converted to excitons.

## 6. Identifying the TR profiles of the $A_W$ resonance induced by various effects

6.1 The electron-transfer TR profile

The electron transfer from MoSe$_2$ to WS$_2$ plays a role similar to electron doping WS$_2$, and hence we can obtain the electron-transfer TR profile by measuring steady RC spectra with varying electron doping in WS$_2$. Specifically, with a large $V_E = 7$ V, we can electron dope WS$_2$ and keep MoSe$_2$ neutral, as shown in Figure S4a,b, in which the reflectance of WS$_2$, instead of MoSe$_2$, is largely modulated by varying $V_D$. The electron-transfer TR profile is given by $(RC_V - RC_{V_0})/(RC_{V_0} + 1)$, as shown in Figure S4c. $RC_V$ is the RC spectrum of electron doped WS$_2$ with $V_D = V$, and $RC_{V_0}$ is the RC spectrum of neutral WS$_2$ with $V_D = V_0 = 1.089$ V. The normalized TR profiles obtained with various dopings have similar shapes (inset of Figure S4c).

We correlate the magnitude of the electron-transfer TR profile with $n_e$ in WS$_2$, by extracting the maximum of the TR profiles as a function of $V_D$, shown in Figure S4d. For $V_D$ above 1.18 V, the maximum shows linear behavior, indicating good ohmic contact between the graphite electrode and TMD monolayers; however, for $V_D$ below 1.18 V where $n_e$ is small, the maximum shows superlinear behavior, which likely arises from non-ohmic contact between graphite electrode and TMD that is commonly seen for slightly doped TMD[5]. We obtained the change rate of the TR-profile maximum per $n_e$ from the linear regime ($V_D > 1.18$ V) by using a parallel-plate capacitor model (see the section of calculation of $n_e$ and $\Delta_C$), which is applied to the whole doping regime.

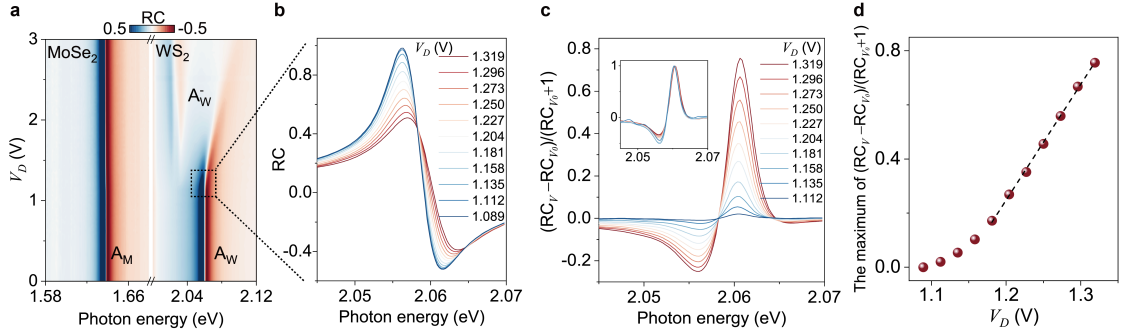

**Figure S4.** Determination of the electron-transfer TR profile. (a) The contour plot of steady RC spectrum as a function of doping in WS2, around the resonances of MoSe2 and WS2. $V_E$ is fixed at 7 V. (b) The linecuts of the RC spectra around the $A_W$ resonance at the initial doping regime, indicated by the dashed box in (a). (c) The contrast between the RC spectrum with electron doped WS2 ($RC_V$) and that with neutral WS2 ($RC_{V_0}$ and $V_0$ =1.089 V), defined as $(RC_V - RC_{V_0})/(RC_{V_0} + 1)$, which is equivalent to the electron-transfer TR profile. The inset shows normalized electron-transfer TR profiles with varying electron doping in WS2. (d) The maximum of the electron-transfer TR profile as a function of $V_D$.

6.2 The temperature-increase TR profile

To obtain the TR profile induced by lattice temperature increase, we measured steady RC spectra of the $A_W$ resonance with increasing sample temperature, as shown in Figure S5a. Both WS2 and MoSe2 are kept at neutral. As the temperature increases from 10 to 31 K, the $A_W$ resonance shows red shift, along with slightly decreased magnitude. The TR profile is given by $(RC_T - RC_{T_0})/(RC_{T_0} + 1)$, shown in Figure S5b. $RC_T$ and $RC_{T_0}$ are the RC spectra at elevated temperature $T$ and base temperature $T_0$ ($\approx$ 10 K), respectively. The maximum of the TR profile increases linearly with temperature increment $\Delta T$ that is defined as $T - T_0$, shown in Figure S5c. The normalized TR profiles obtained with various $\Delta T$ have similar shapes, except slight red shift at higher $T$ (the inset of Figure S5b).

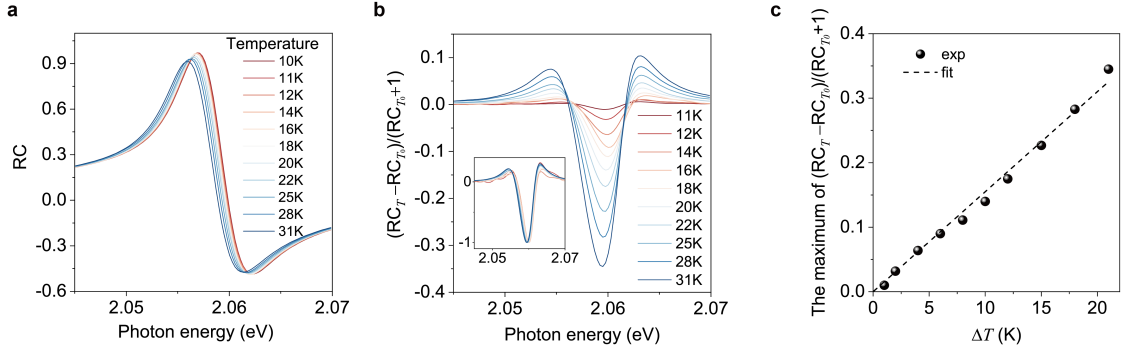

**Figure S5.** Determination of the temperature-increase TR profile. (a) The steady RC spectra of the $A_W$ resonance of WS$_2$ at various temperatures. (b) The contrast between the RC spectrum at elevated temperature ($RC_T$) and that at base temperature ($RC_{T_0}$ and $T_0$ =10 K), defined as $(RC_T - RC_{T_0})/(RC_{T_0} + 1)$, which is equivalent to the temperature-increase TR profile. The inset shows the normalized TR profiles. (c) The maximums of the temperature-increase TR profiles as a function of temperature increment.

6.3 The screening TR profile

The mobile photoexcited excitons in MoSe$_2$ will enhance the surrounding screening on WS$_2$ and hence modulate the reflectance resonances of WS$_2$. To mimic the screening-induced TR profile, we enhanced the surrounding screening on WS$_2$ by electron doping MoSe$_2$, and measured steady RC spectra, as shown in Figure S6a. As $n_e$ in MoSe$_2$ increases, the $A_W$ resonance exhibits a red shift, along with slight reduced magnitude, shown in Figure S6b. The TR profile is given by $(RC_{n_e} - RC_0)/(RC_0 + 1)$, where $RC_{n_e}$ and $RC_0$ are the RC spectra of the $A_W$ resonance with electron doped and neutral MoSe$_2$, respectively, as shown in Figure S6c. The maximum of the TR profile shows a quick saturation behavior as a function of $n_e$ (Figure S6d). The normalized TR profiles obtained with various $n_e$ have similar shapes, except slight red shift for larger $n_e$ (the inset of Figure S6c).

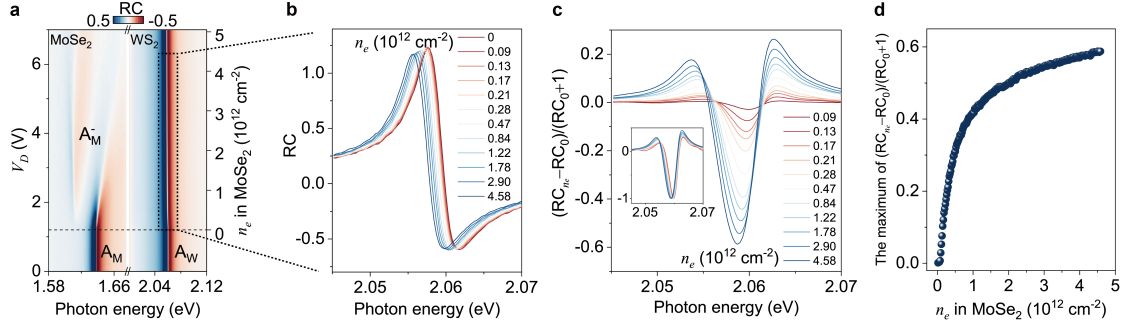

**Figure S6.** Determination of the screening TR profile. (a) The contour plot of steady RC spectrum as a function of doping in MoSe2, around the resonances of MoSe2 and WS2. $V_E$ is fixed at 0 V. The dashed box denotes the electron-doping regime of MoSe2. (b) The linecuts of the RC spectra around $A_W$ resonance at varying $n_e$, indicated by the dashed box in (a). (c) The contrast between the RC spectra of $A_W$ resonance with electron doped MoSe2 ($RC_{n_e}$) and that with neutral MoSe2 ($RC_0$), defined as $(RC_{n_e} - RC_0)/(RC_0 + 1)$, which is equivalent to the screening TR profile. The inset shows normalized TR profiles with varying electron doping in MoSe2. (d) The maximums of the screening TR profiles as a function of $n_e$ in MoSe2.

6.4 Discussion of the three effects and the two-profile analysis

Let's first compare the electron-transfer TR profile with $n_e$=3.5×10$^{10}$ cm$^{-2}$ in WS2, the temperature-increase TR profile with $\Delta T$=21 K, and the screening TR profile with $n_e$=0.47 ×10$^{12}$ cm$^{-2}$ in MoSe2, shown Figure S7a-c, respectively. Their normalized profiles are shown in Figure S7d. We can see that, the electron-transfer TR profile is distinctly different from the other two almost identical TR profiles.

Although the temperature-increase and screening TR profiles cannot be spectrally distinguished, we expect that, the temperature-increase should lead to a larger contribution to the experimental TR signal than that of the screening effect. The reason is that, the screening-TR profile calibrated in the steady measurement exaggerates the screening ability of photo-excited carriers. In the steady measurement, the screening effect on excitons of WS2 origins from the doped electrons in MoSe2, which obeys the 1/$r^2$ Coulomb potential between electron and exciton, where $r$ denotes the electron-exciton separation. While in the pump-probe measurement, the screening effect comes from the photo-excited excitons, which obeys the 1/$r^3$ Coulomb potential of dipole-dipole interaction and exhibits a faster decay with $r$ than the steady case. Moreover, owing to the inserted 3L-hBN between MoSe2 and WS2 which enlarges the distance and weakens the dipole-dipole interaction, the screening effect is excluded in the analysis.

We use the electron-transfer TR profile of $n_e$=3.5×10$^{10}$ cm$^{-2}$ and the temperature-increase TR profile of $\Delta T$ =21 K as the two bases to fit the experimental TR spectra, termed as the two-profile analysis. For each fitting, the two TR profiles are rescaled and spectrally shifted within a narrow range about 1 nm, which are optimized based on the principle of least squares; the maximums of the rescaled TR profiles are used to extract $n_{ET}$ and $\Delta T^*$, according to the relations obtained in Figure S4d and Figure S5c, respectively. Noteworthy, choosing another set of the TR profiles with different $n_e$ or $\Delta T$ will not alter the results, as the normalized TR profiles with varying either $n_e$ or $\Delta T$ have similar shapes (Figure S4c and Figure S5b).

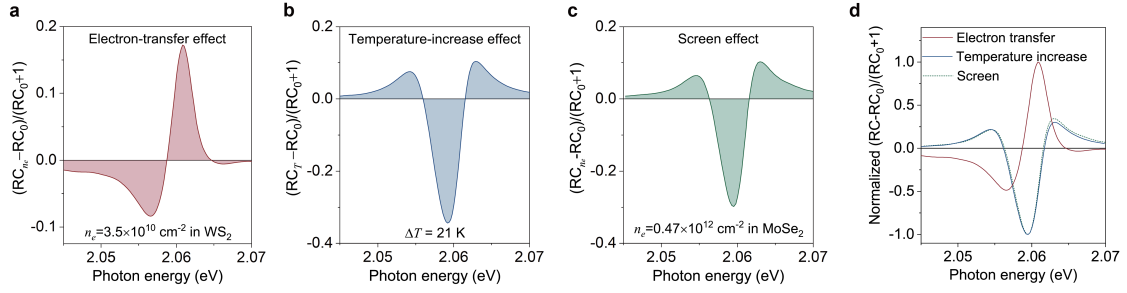

**Figure S7.** Comparing the TR profiles of $A_W$ resonance induced by various effects. (a) The electron-transfer TR profile with $n_e = 3.5 \times 10^{10}$ cm$^{-2}$ in WS$_2$. (b) The temperature-increase TR profile with $\Delta T = 21$ K. (c) The screening TR profile effect with $n_e = 0.47 \times 10^{12}$ cm$^{-2}$ in MoSe$_2$. (d) The normalized TR profiles of the three effects.

## 7. The two-profile analysis of the TR spectrum

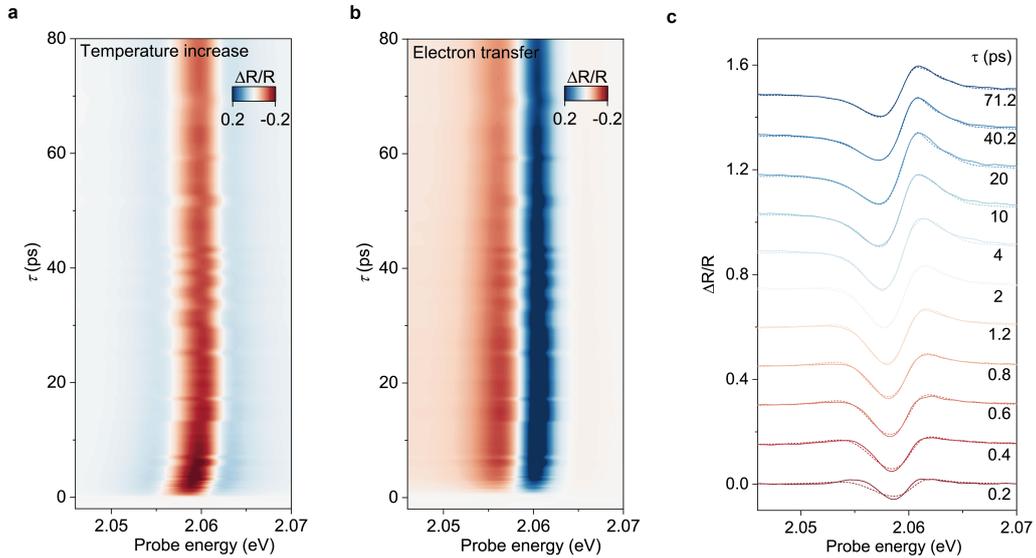

**Figure S8.** The two-profile analysis of the TR spectrum. (a) and (b) are respectively the contour plots of the temperature-increase and electron-transfer TR profiles as a function of $\tau$, obtained by fitting the TR spectra in Figure 2 with the two-profile analysis. (c) The TR spectra (solid) and fitting results (dashed) at varying $\tau$. The curves are displaced vertically for clarity.

## 8. $\Delta T^*$ dynamics with varying doping in MoSe$_2$.

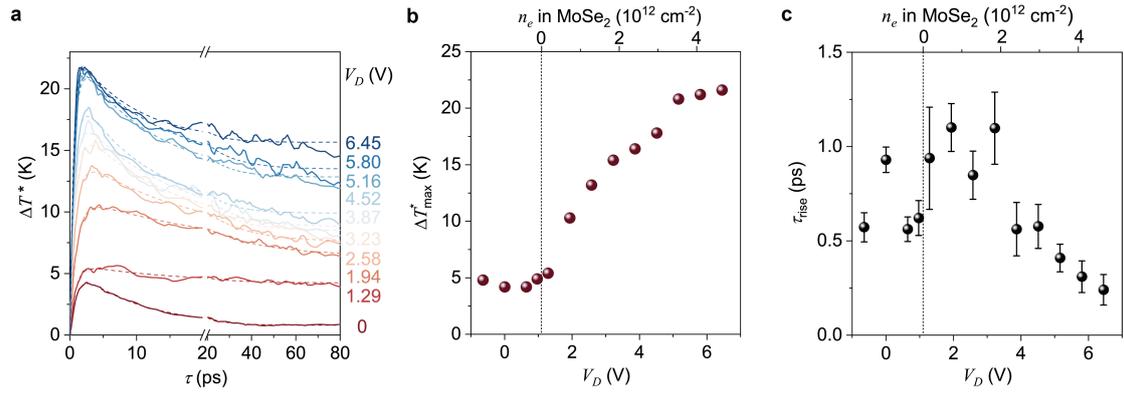

**Figure S9.** $\Delta T^*$ dynamics with varying doping in MoSe$_2$. (a) The doping dependence of $\Delta T^*$ dynamics, which are extracted from the TR data in Figure 3. $V_D$ varies from 0 V to 6.45 V. $V_E$ is fixed at 0 V. The dashed line is a two-exponential fitting. (b) The maximum of $\Delta T^*$ dynamics, labelled as $\Delta T^*_{max}$, as a function of $V_D$. (c) The rise time constant of $\Delta T^*$ dynamics as a function of $V_D$.

## 9. The TR spectrum around the $A_M$ resonance upon electron doping MoSe$_2$.

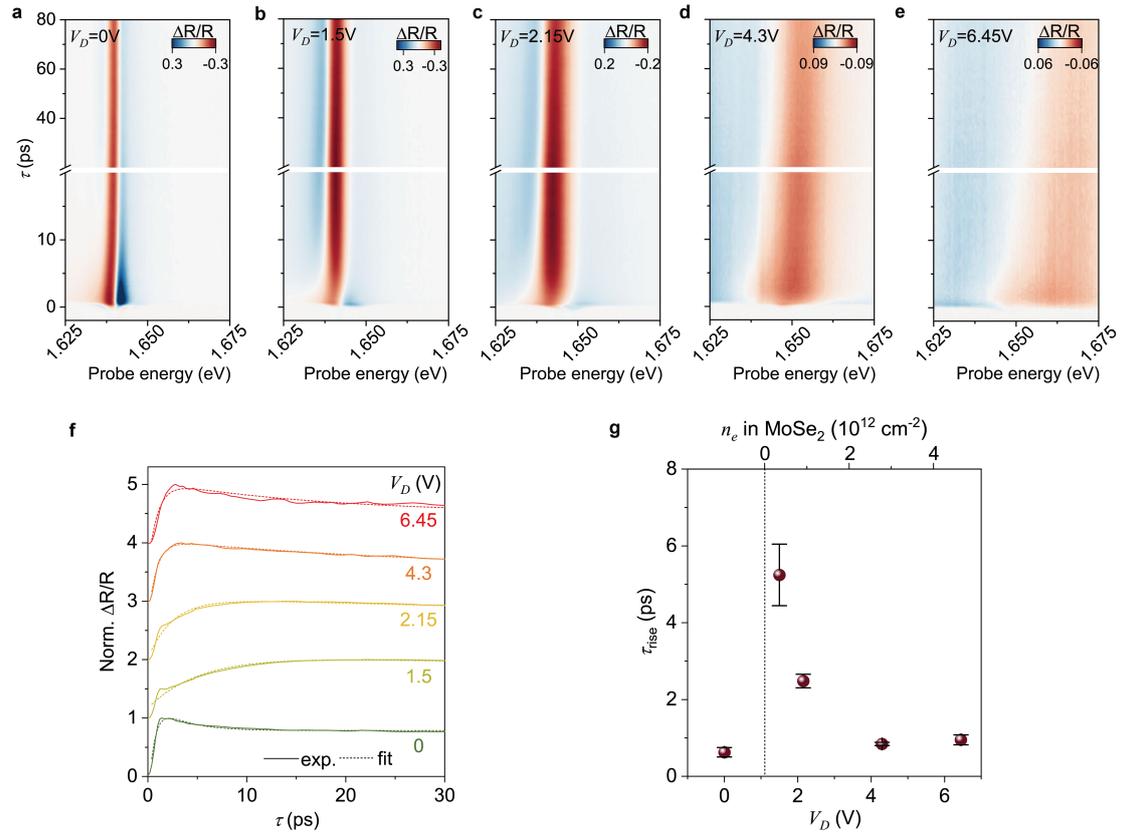

**Figure S10.** The TR spectrum around the $A_M$ resonance of MoSe$_2$ upon electron doping MoSe$_2$. (a-e) Contour plots of the TR spectrum around the $A_M$ resonance as a function of $\tau$, with $V_D$ varying from 0 V to 6.45 V while $V_E$ fixed at 0 V. (f) The normalized dynamics of the $A_M$ resonance, which are extracted from the dip magnitude of the TR spectrum. The dashed line denotes a two-exponential fitting. (g) The rise time constant of the $A_M$ resonance as a function of $V_D$, obtained from the fitting in (f).

## 10. The pump-fluence dependence of $\Delta T^*$ dynamics.

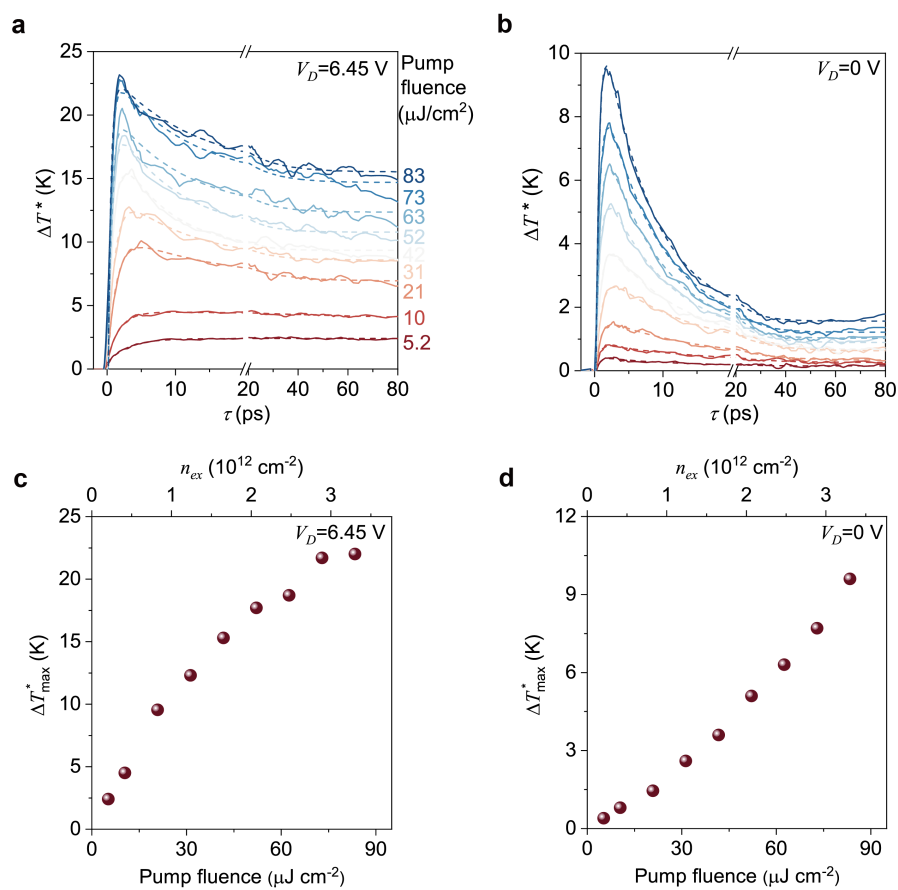

**Figure S11.** The pump-fluence dependence of $\Delta T^*$ dynamics. (a,b) $\Delta T^*$ dynamics with varying pump fluence with $V_D$ = 6.45 and 0 V, which are extracted from the TR data in Figure 4. $V_E$ is fixed at 0 V. (c,d) $\Delta T^*_{max}$ as a function of pump fluence.

## 11. The electric-field dependence of $\Delta T^*$ dynamics.

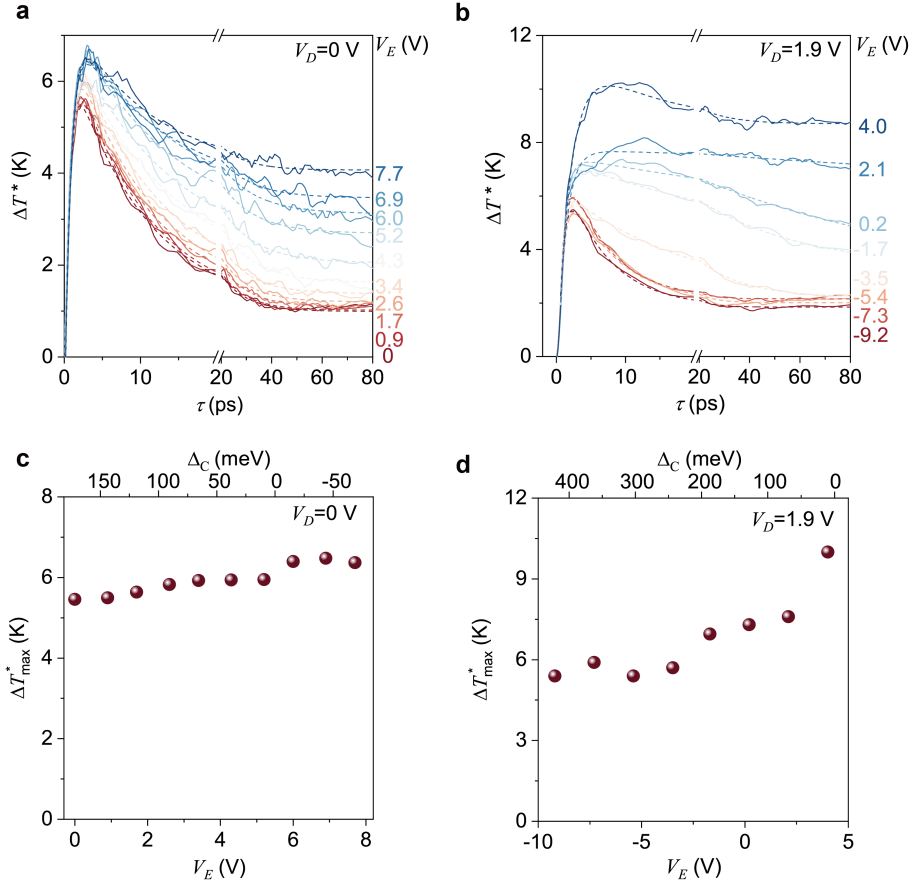

**Figure S12.** The electric-field dependence of $\Delta T^*$ dynamics. (a,b) $\Delta T^*$ dynamics for varying $V_E$ upon $V_D$=1.9 V and 0 V, which are extracted from the TR data in Figure 5. (c,d) $\Delta T^*_{max}$ as a function of $V_E$.

## 12. Discussion of hot-electron extraction at ambient condition

We think the doping-enhanced hot-electron extraction found at 10 K can be applied to 300 K, which can be understood as follows.

The excess energy of hot excitons, defined as the energy difference between the pump photon and conduction-band edge, is about 100 meV in our experiment, which will lead to the effective electronic temperature much larger than the thermal energy at 300 K (about 25 meV). In both scenarios of 10 K and 300 K, besides exciton-carrier interactions, the excess energy of the hot excitons is lost to the lattice through optical phonon emission[6]. The probability of optical phonon-emission is proportional to $(n + 1)$, where $n$ obeys the Bose-Einstein distribution, $1/(e^{\frac{\hbar\omega}{k_B T}} - 1)$. $k_B$ is Boltzmann's constant, $T$ is the lattice temperature, and $\hbar\omega$ is the optical phonon energy. With temperature increasing from 10 K to 300 K and using $\hbar\omega$ of about 20 meV[7], $(n + 1)$ only varies from 1 to 1.8. Thus, in spite of some enhanced optical phonon emission, we expect that the competition between the exciton-electron scatterings and the exciton-phonon scattering should be similar for both 10 K and 300 K, and hence the hot-electron extraction driven by exciton-electron interactions should still work at ambient conditions.

## 13. Supporting references